\documentclass[prl,twocolumn,showpacs,preprintnumbers,amsmath,amssymb]{revtex4}

\usepackage{amsmath}
\usepackage{amssymb}
\usepackage{graphicx}
\usepackage{dcolumn}
\usepackage{bm}
\usepackage{psfrag}
\usepackage{rotating}
\usepackage{color}

\newcommand{\txr}[1]{\textrm{#1}}

\begin{document}


\title{Quasi-deterministic transport of Brownian particles in an
  oscillating periodic potential}
\author{Pawel Romanczuk} \email{romanczuk@physik.hu-berlin.de}
\author{Felix M{\"u}ller} \author{Lutz Schimansky-Geier}
\affiliation{%
  Institute of Physics, Humboldt University Berlin, Newtonstr. 15,
  12489 Berlin, Germany }%
\begin{abstract}
  We consider overdamped Brownian dynamics in a periodic potential
  with temporally oscillating amplitude. We analyze the transport
  which shows effective diffusion enhanced by the oscillations and
  derive approximate expressions for the diffusion coefficient.
  Furthermore we analyze the effect of the oscillating potential on
  the transport if additionally a constant force is applied. We show
  the existence of synchronization regimes at which the deterministic
  dynamics is in resonance with the potential oscillations giving rise
  to transport with extremely low dispersion. We distinguish slow and
  fast oscillatory driving and give analytical expressions for the
  mean velocity and effective diffusion.
\end{abstract}

\maketitle

In recent decades there has been intense research on transport
phenomena in Brownian dynamics. Numerous publications analyze the diffusive
transport in spatially periodic potentials and report interesting
behavior of the diffusion coefficient
\cite{reimann_giant_2001,lindenberg_dispersionless_2007, borromeo_resonant_2008}. Diverse sorts of ratchets have been
investigated
\cite{reimann_brownian_2002,anishchenko_nonlinear_2002,hanggi_marchesoni_rmp}
where particles move in periodic potentials and time dependent forces
or modulations of the potential enhance a directed flow. Most of the
research focused on mean drift within such systems and on the
conditions under which the directed transport is maximized. Also the
diffusion coefficient in these temporally changing potentials have
been calculated \cite{freund_diffusion_1999} and it was proposed to
take the diffusion coefficient to evaluate the precision of stochastic
directed transport \cite{lindner_noise-induced_2002}.

In this spirit we will focus this work on overdamped Brownian
transport in a temporally oscillating and spatially periodic
potential. The interplay of the different time scales in the system,
given by the period of the oscillating potential, the relaxation time
of the deterministic dynamics and the diffusion time gives rise to
non-trivial dynamical phenomena, such as an oscillation driven
enhancement of the effective diffusion. Otherwise if
additionally external forces are applied, the particle's motion
becomes quasi-deterministic following the oscillations in
direction of the applied force jumping several periods of the potential
with minimal diffusion. 
Experimental systems where such temporal modulations could be realized
are free-flow dielectrophoresis \cite{ajdari_free-flow_1991},
colloidal particles in optical fields
\cite{faucheux_optical_1995,tatarkova_brownian_2003,bleil_directing_2007},
Josephson-junctions \cite{kautz_noise_1996,sterck_rectification_2009}
or paramagnetic colloids in magnetic fields
\cite{tierno_transport_2008,auge_magnetic_2009}. With small
modification the dynamics can also describe neuronal
activity being one type of a firing theta-neuron
\cite{kopell_ermentrout_86,neiman_russell_05}.

We consider an overdamped Brownian particle under the influence of thermal fluctuations in spatially periodic potential with
harmonically oscillating amplitudes.  The corresponding Langevin
equation written in dimensionless form reads
\begin{equation}\label{dimless}
  \dot{y}  =\cos (y) \sin (\Omega\tau)+F+\sqrt{2 D}\, \xi(t), 
\end{equation}
with driving frequency $\Omega$, constant force $F$, and noise
intensity $D$.

As characteristic observables we investigate the asymptotic drift
velocity and the effective diffusion coefficient
\begin{eqnarray}
  && v_\txr{drift}=\lim_{\tau \to\infty}\frac{\langle y(\tau) \rangle -\langle y(0) \rangle}{\tau}\\
  &&D_{\txr{eff}} = \lim_{\tau \to \infty} \frac{1}{2}\frac{d}{d \tau}\langle (y-\langle y(\tau)\rangle)^2 \rangle .
\end{eqnarray}
Here $\langle \cdot \rangle$ denotes the ensemble average.

In the case of non-vanishing flux in the system ($F\neq 0$) we measure
the quality of the directed transport by the so called P\'eclet number
\cite{freund_diffusion_1999}: $\txr{Pe}=v_\txr{drift} L/D_\txr{eff}$
where $L$ is the characteristic length scale of the system. Here we set
$L$ equal to the spatial wave length $\lambda=2\pi$.  For $\txr{Pe}<1$
diffusion dominates the dynamics, and the directed transport plays a
minor role in comparison with the non-directed spread of the
probability distribution. For $\txr{Pe}>1$ the transport is dominated
by the drift. The limit of $\txr{Pe}\to\infty$ corresponds to a
deterministic transport with vanishing effective diffusion on the
characteristic length scale $L$.

First we consider the case without bias in the system for $F=0$.  The
spatial and temporal symmetry in the potential prevents any directed
flux within the system ($v_\txr{drift}=0$). For large noise
intensities the influence of the potential is negligible and the
dynamics corresponds to a free Brownian motion with $D_{\txr{eff}} =
D$ whereas in the limit of vanishing noise the interplay of two time scales
controls the dynamics. On the one hand the external driving period
$T=2\pi/\Omega$ and secondly the intrinsic relaxation time $\tau_{r}$
from an unstable potential maximum to a stable minimum.

At oscillation frequencies much faster than the relaxation time
$\Omega \gg 1/\tau_r$ the potential is self-averaged to be effectively
flat for the particle dynamics and the effective diffusion also
converges to the free Brownian diffusion coefficient $D$ for $\Omega
\to \infty$.  However, for a relaxation time shorter than the half
oscillation period $\tau_r <T/2=\pi/\Omega$ trajectories are able to
approach the next minimum until the oscillation changes the minimum
into a metastable maximum. Involving a small amount of noise the
particle performs effectively discrete jumps between the minima of the
potential for the extreme potential settings (maximal barriers).

This behavior can be easily described by an oscillation induced random
walk.  The equally distributed jumps to the left and right take place
at discrete time intervals $T/2$ and correspond to the switching of the
fixed points (two jumps per full period). The jump length is given by
$l=\lambda/2 = \pi$.  Thus the
probability to find the particle at the position $y_i=i\cdot l$ after
$N$ jumps is given by the binomial distribution, which converges
towards a Gaussian distribution in the long time limit describing
diffusion with effective coefficient $D_{\txr{eff}}=\pi\Omega/2$.
As discussed above for large frequencies the effective diffusion will
reach asymptotically the noise intensity $D$. So the linear relation
can not hold for the whole frequency range but has to pass a maximum.

\begin{figure}
  \includegraphics[width=0.9\linewidth]{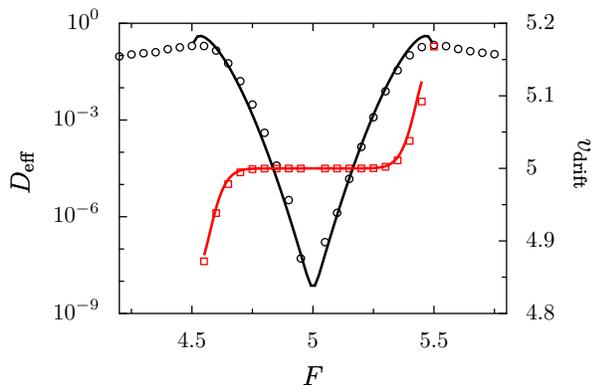}
  \caption{ Simulation of Eq. \ref{dimless} at the 1:1 synchronization
    ($\Omega=5.0$, $D=0.05$): $D_\txr{eff}$ (circles) and the
    $v_\txr{drift}$ (squares). The solid lines are the corresponding
    analytical results for the mean velocity (Stratonovich
    \cite{stratonovich_topics_1967}) and the effective diffusion from
    Eq. \ref{Deff_kramers}.}
\label{fig_deff2}
\end{figure}

An initial Gaussian distribution around a stable fixed point
(potential minimum) $y_n$ splits as the fixed point becomes unstable
(transforms into a potential maximum). For intermediate frequencies
only a certain fraction of the particle ensembles can reach the next
minimum. The remaining part moves back into the initial position.  We
attempt to describe this mechanism by approximating the probability
$u(\Omega,D)$ of a particle starting at $y_n+\delta$ ($\delta\geq0$)
from which a particle is able to arrive at the neighboring fixed point
$y_{\txr{end}}=y_{n+1}-\epsilon$. Here $\epsilon$ represents a small
distance which the particle overcomes by fluctuations alone.  This
probability is given by the complementary error function
\begin{equation}
	\label{erfc}
	u(\Omega,D)=\txr{erfc}\left(\frac{\delta}{\sqrt{2\sigma^2}}\right),
\end{equation}
with the width $\sigma^2$ as the particle distribution at $y_n$ and
$\delta = 2\, \txr{arctan}\left[ \exp \left(-2 / \Omega \right)
  \txr{cot} \left( \epsilon/2 \right) \right]$ as the initial
deviation.  The new expression for the effective diffusion coefficient
has to be scaled by taking Eq.~\ref{erfc} into account and we get
\begin{equation}
	\label{Deff2}
	D_{\txr{eff}}=\frac{\pi \Omega}{2}\txr{erfc}\left(\sqrt{\frac{2}{\sigma^2}}\, \txr{arctan}\left[ e^{-\frac{2}{\Omega}} \txr{cot} \left( \frac{\epsilon}{2} \right)  \right]\right).
\end{equation}
\begin{figure}
  \includegraphics[width=0.9\linewidth]{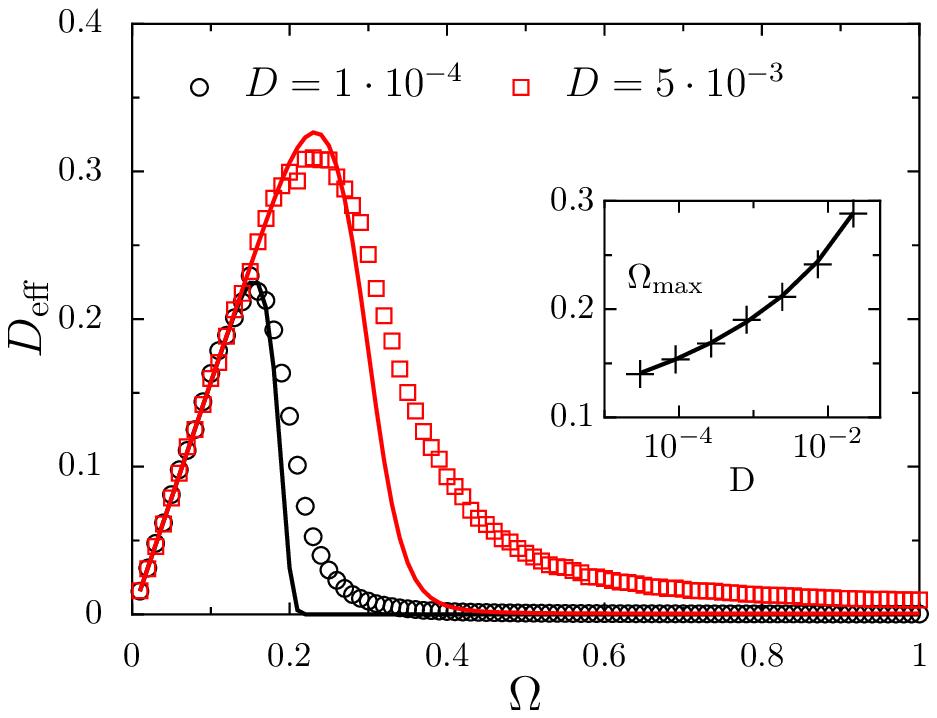}
  \caption{Effective diffusion vs. driving frequency from simulations
    for $F=0$ and $D=1.0 \cdot 10^{-4}$ (circles), $D=5.0 \cdot
    10^{-3}$ (squares) in comparison with results of of Eq.
    \ref{Deff2} (solid lines) for $e=0.09$ and $b=150$. The inset
    shows the prediction on the the position $\Omega_\txr{max}$ of the
    maximum of $D_\txr{eff}$ vs. $D$ (solid line) compared to
    numerical results.}
\end{figure}
The parameters $\epsilon$ and $\sigma^2$ remain unknown. As $\epsilon$
represents a minimal length and $\sigma^2$ is the variance of the
initial Gaussian distribution, we make the following ansatz for their
dependence on $D$: $\sigma^2=bD$ and $\epsilon=e\sqrt{D}$, where
$e$, $b$ are undetermined constants.  This result is compared with
numerical simulations. Examples are shown in Fig.~\ref{fig_deff2}.  A
reasonable choice of the two undetermined constants is $e=0.09$ and
$b=150$. The linear increase of $D_\txr{eff}$ is recovered together
with a maximum at $\Omega_\txr{max}$ in agreement with the numerics
(circles and squares in Fig.~\ref{fig_deff2}). The inset in
Fig.~\ref{fig_deff2} shows the dependence of the position
$\Omega_\txr{max}$ of the effective diffusion over the noise intensity
$D$ over three orders of magnitude. It illustrates that a single
choice of $e$ and $b$ allows predictions on the impact of noise on
$\Omega_\txr{max}$.  Please note that the approximation is not valid
at very large frequency values as 
$\lim_{\Omega\to \infty}D_\txr{eff}(\Omega,D)$ diverges.

The addition of a constant force leads to a temporally constant tilt
of the oscillating periodic potential.  In this case a critical force
can be defined, where for all times no potential barriers obstruct the
drift motion of the particle: $F_\txr{crit}=1$.  In contrast to a
static washboard potential we observe a finite drift for vanishing
noise $D\to0$ also at subcritical forces $F<F_\txr{crit}$ due to the
oscillations of the potential.

\begin{figure}
  \begin{center}
    \includegraphics[width=0.90\linewidth]{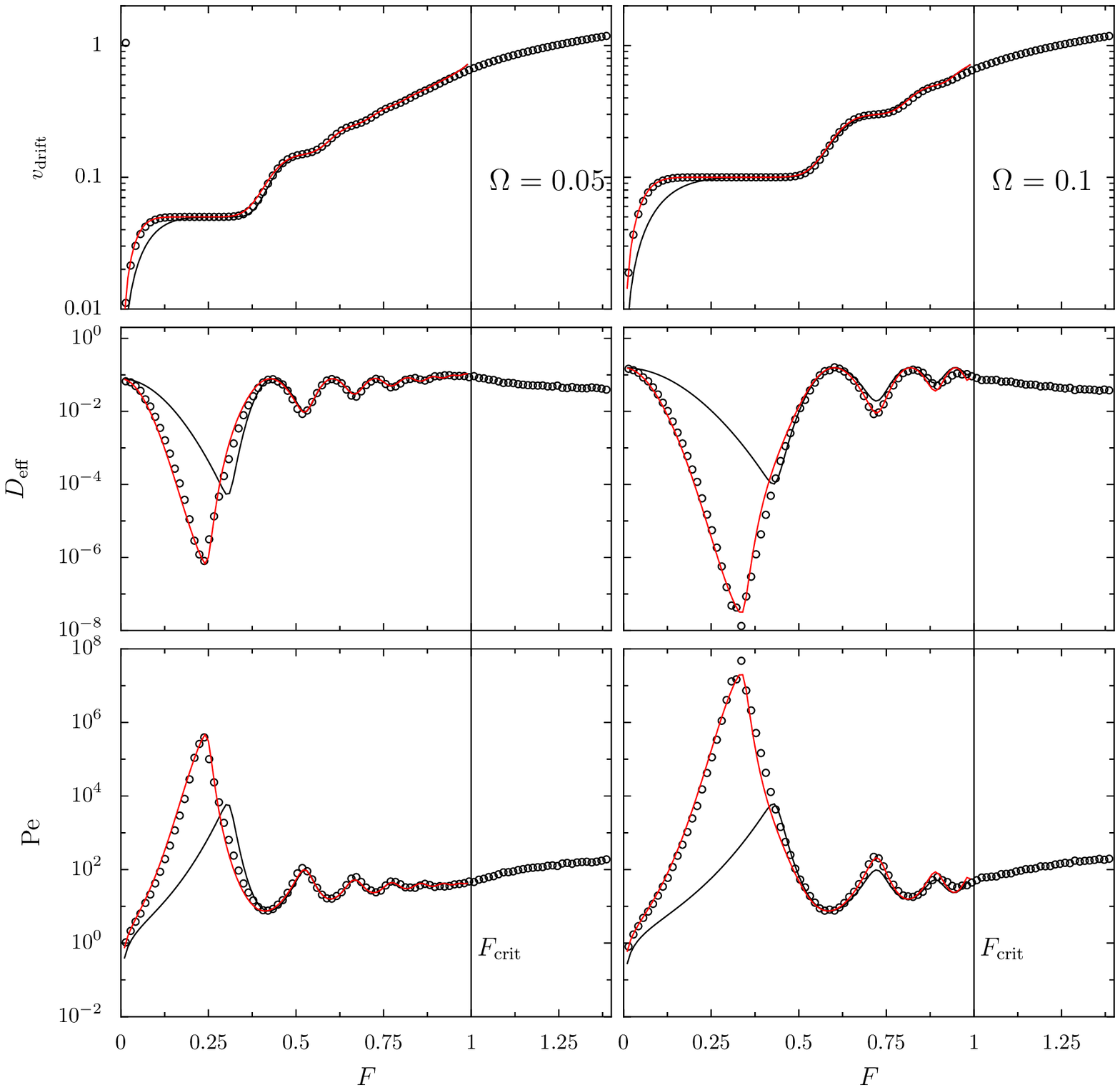}
    \caption{Comparison of simulations (symbols) with the theoretical
      results at low $\Omega$ (solid lines): mean velocity
      $v_\txr{drift}$, effective diffusion $D_\txr{eff}$ and P\'eclet
      number Pe vs. bias $F$ for $\Omega=0.05$ (left column) and
      $\Omega=0.1$ (right column). The black lines show the linear
      ansatz $G=a F$ with $a=0.8$, the red lines show the non-linear
      ansatz with $G=a(F)F$ with $a_1=1.6$, $a_2=0.8$, $c=10$ for both
      frequencies. The only changing parameter is $F_t$: for
      $\Omega=0.05$ $F_t=0.28$, for $\Omega=0.1$ $F_t=0.4$. For both
      examples shown $b=0.08$.\label{fig_peclet}}
		\end{center}
\end{figure}

The drift speed $v_\txr{drift}$ averaged over initial conditions
$y_\txr{ini}\in(0,2\pi)$ as a function of the $F$ shows different
plateaus at $v_\txr{drift}=\Omega, 3\Omega, 5\Omega,\dots$,
corresponding to 1:1, 1:3 and 1:5 synchronizations (Figs.
\ref{fig_deff2},\ref{fig_peclet}). This behavior
resembles characteristics of a driven oscillator, where such plateaus
correspond to entrainment regimes of the oscillator to the external
driving. Furthermore externally driven stochastic oscillators
show a strong inhibition of the effective phase diffusion in the
synchronized state \cite{anishchenko_nonlinear_2002}, which is also
observed in our system.

Based on these similarities we attempt to describe the dynamics of our
system in vicinity of the synchronization regime by a corresponding
ansatz. We change into the co-moving frame $z(t)=y(t)-\Omega t$ and
consider the averaged deterministic dynamics over one oscillation
period $T=2\pi/\Omega$.
Due to the undetermined time-dependence of $z(t)$ we are not able to
obtain a general solution of the averaged dynamics. We may write $z$
as a Taylor series $z(t') = z(0)+z'(0)t'+\dots$ with $t'=t/T$. Under
the assumption $z(t')$ changes slowly over one oscillation period we
keep only the $0$-th order and set $z(t')=z_T$. With this
approximation we obtain the stochastic Adler equation:
\begin{equation}\label{avgz2}
\dot z_T  = \Delta - \frac{1}{2}\sin z_T + \sqrt{2 D}\xi(t)
\end{equation}
with $\Delta =| F - \Omega |$. 

Please note that due to the assumptions the equation holds only in the limit of fast oscillations with respect to the intrinsic relaxation time $\Omega\gg\tau_r^{-1}$. Thus at 1:1 synchronization ($F=\Omega$, $\Omega\gg1$) we are in fact in the super-critical force regime. In the co-moving frame even for $F>F_\txr{crit}=1$  the reduced Equation \ref{avgz2} describes a particle moving in a tilted stationary periodic potential.

An analytical solution for $v_\txr{drift}$ close to synchronization
can be taken directly from the classical work of Stratonovich on the
mean phase of an entrained stochastic oscillator
\cite{stratonovich_topics_1967}, whereas $D_\txr{eff}$ can be
calculated in the case of small noise $D\ll1/2$, from the Kramers
rates $k^\pm$ to neighboring minima
of the co-moving stationary potential $D_{\txr{eff}} = \frac{\lambda^2}{2} \left( k^+ + k^- \right)$:
\begin{align}
  \label{Deff_kramers}
  D_{\txr{eff}} = \frac{w_0}{2\pi}\, \txr{cosh}\left(
    \frac{\Delta}{D}\pi \right) e^{-\frac{2}{D} ( \Delta
    \txr{arcsin} (2\Delta) + w_0 )},
\end{align}
with $w_0 = \frac{1}{2}\sqrt{1-4\Delta^2}$.

The breakdown of effective diffusion in the synchronization regime
correspond to giant P\'eclet numbers, which indicate a
quasi-deterministic Brownian transport. For low noise strength
($D<0.01$) and an optimal choice of $F$ the system shows
no dispersion even for extremely large times
($t>10^5 T$, $\sim10^4$ particles).

At low $\Omega$ ($<1$) the first synchronization regimes are located
at subcritical forces $F<1$. The dynamics can be described as a
\emph{slip} and \emph{stick} motion dominated by the oscillating
barriers.  We assume a $\delta$-peaked probability distribution at the
begin of the sliding motion (\emph{slip} phase) at the position $y_0$
corresponding to the fixed point where the particle is located at the
begin of the step. As soon as temporal barriers occur, the particle
relaxes at $y_{l}=l\lambda/2$ with $l=(2n+1)$ and $n\in \mathbb N$
(\emph{stick} phase).  The envelope of the $\delta$-like distributions
of the relaxed particles is assumed to be Gaussian with a mean $\mu_e$
and width $S$.  The probability to find the particle at the unstable
fixed points $y_m$ with $m =2n$ at the end of the step vanishes.

The total probability to find the particle close to fixed point
$y_{l}$ at the end of the step reads
\begin{align}
  w_l = \frac{1}{2}\left[\txr{erf}\left(\frac{y-\mu_e}{\sqrt{2 S\Delta
          \tau}}\right)\right]_{(l-1)\lambda/2}^{(l+1)\lambda/2}
  \label{th_prob}.
\end{align}
Thus we can formally calculate the mean position and the variance
after a single step
\begin{align}
\langle y_e\rangle & = \sum_{n=-\infty}^{\infty} y_{2n+1} w_{2n+1}, \label{th_mean} \\  
\langle (y- \langle y_e \rangle)^2 \rangle  & = \sum_{n=-\infty}^{\infty} \left(y_{2n+1}-\langle y_e \rangle\right)^2 w_{2n+1} \label{th_var},
\end{align}
with $y_{2n+1}=(n+\frac{1}{2})\lambda$.  As the distribution $P_e(y)$
has a finite width and most of the probability will be concentrated
within few minima of the potential around the mean $\mu_e$. It is 
sufficient to consider only a finite number of points $n\in
[n_{min},n_{max}]$ around $\mu_e$ and calculate the mean position and
the variance by renormalizing the probabilities accordingly $\tilde
w_{2n+1}= w_{2n+1}/\sum_{n=n_{min}}^{n_{max}} w_{2n+1}$. With results
obtained in Eqs.~\ref{th_mean} and \ref{th_var} we can calculate the
mean drift and the effective diffusion as
\begin{equation}
  v_\txr{drift} = 2\langle y_e \rangle/ T,\,~~~D_\txr{eff} = \langle (y- \langle y_e \rangle)^2
  \rangle/T \label{th_Deff}\,.
\end{equation}

The mean of the particle probability distribution moves in the
direction of the bias with an effective force $G=G(F,\Omega)$ within a
time interval $\Delta \tau(F,\Omega)$ caused by the combined effect of
the constant bias $F$ and the oscillating potential.  The time of
vanishing barriers is $\Delta \tau=2\arcsin(F)/\Omega$ and for $G$ we
choose a linear dependence on $F$: $G(F)=a F$. Averaging over the
potential oscillations leads to $a=1$.

The diffusion coefficient $S$ in Eq. \ref{th_prob} depends on the
applied force and shows a large increase close to the critical force
$F_\txr{crit}$ likewise reported in \cite{reimann_giant_2001}. For
simplicity we assume here a linear increase $S=b F$.  The simplified
theory given in Eqs. \ref{th_prob}, \ref{th_Deff} is compared with
simulations of the full system for $\Omega=0.03,0.05$ in
Fig.~\ref{fig_peclet}.  Symbols represent the numerical results while
the black solid line represents the analytical calculation for $a =
0.8$ and $b=0.04$. The multiple peaks of the P\'eclet number with
decreasing magnitude are reproduced together with the shift of the
peaks towards larger $F$ for increasing $\Omega$ which leads to a
decrease in the number of peaks for $F<F_\txr{crit}$.  However the
most obvious difference is the position and the height of the first
peak of the P\'eclet number.

A small bias $F$ is potentiated by the slope of the oscillating
potential therefore the parameter $a$ should be chosen as $a>1$. We describe
this through a step function around $F_t$ as :$a(F)=a_2+(a_1-a_2)\left[
  \txr{tanh}\left(c (F_t-F)\right)-1\right]/2$ This
phenomenological modification of the model leads to the red (gray)
line and shows agreement to the numerical simulations. Please note that $F_t$
is the only changing parameter 
in calculations of the non-linear ansatz for different frequencies.
This good agreement of the theoretical results with the numerical
simulations shows that the theory accounts for the decisive mechanism
responsible for the observed behavior: the combination of transport
(\emph{slip} phase) with repeated confinement (\emph{stick} phase) at
temporal minimum of the potential.

In this work we have analyzed the impact of a spatio-temporal
oscillating potential on Brownian transport. We have demonstrated how
the oscillating potential may enhance or suppress the effective
diffusion. We derived an expression for $D_\txr{eff}(\Omega,D)$ for 
vanishing bias $F=0$. We have obtained an optimal driving frequency
which maximizes the diffusive transport. For finite bias $F>0$ we have
shown the occurrence of synchronization regimes, with strongly
suppressed effective diffusion and quasi-deterministic transport at
finite noise strengths revealed through giant P\'eclet numbers.

Via a transformation to the co-moving frame at fast oscillations
analytical expression based on previous results by Stratonovich and
Kramers are found for the effective diffusion and mean velocity. At
low frequencies we describe the system dynamics through a simplified
ansatz of the transport process, which for reasonable choice of
effective parameters are in agreement with the numerical results for
$0<F<F_\txr{crit}$. Our results suggest that in the unbiased case the
introduction of an spatio-temporal periodic force enhance the
diffusive transport, whereas for a finite bias it may significantly
improve the quality of directed Brownian transport by suppressing thermal fluctuations.

The authors thank M. Kostur for the introduction to computing with
CUDA \cite{CUDAGuide}, which allowed a remarkable simulation speed up \cite{januszewski_accelerating_2010}. We acknowledge the
financial support by the DFG through Sfb555.

\end{document}